\documentclass[aps,prl,superscriptaddress,twocolumn]{revtex4-2}

\usepackage{graphics}
\usepackage{graphicx} 
\usepackage{dcolumn}
\usepackage{bm}
\usepackage{epstopdf}
\usepackage[usenames, dvipsnames]{color}
\usepackage{amsmath}

\begin{document}

\title{Magnetic order in the two-dimensional metal-organic framework manganese pyrazinecarboxylate with Mn-Mn dimers}



\author{S.~Calder}
\email{caldersa@ornl.gov}
\affiliation{Neutron Scattering Division, Oak Ridge National Laboratory, Oak Ridge, Tennessee 37831, USA.}

\author{R.~Baral}
\affiliation{Neutron Scattering Division, Oak Ridge National Laboratory, Oak Ridge, Tennessee 37831, USA.}

\author{N.~Narayanan}
\affiliation{University of Missouri Research Reactor (MURR), University of Missouri, Columbia, MO 65211, USA}
\affiliation{Department of Chemistry, University of Missouri, Columbia, Missouri 65211, USA}

\author{L.~D.~Sanjeewa}
\affiliation{University of Missouri Research Reactor (MURR), University of Missouri, Columbia, MO 65211, USA}
\affiliation{Department of Chemistry, University of Missouri, Columbia, Missouri 65211, USA}

\begin{abstract}
The magnetic properties of [Mn(pyrazinecarboxylate)$_2$]$_n$ (Mn-pyrazine), empirical formula C$_{10}$H$_6$MnN$_4$O$_4$,  are investigated  through susceptibility, heat capacity and neutron scattering measurements. The structure of Mn-pyrazine consists of Mn-Mn dimers linked on a distorted 2D hexagonal structure. The weak out of plane interactions create a quasi-2D magnetic material within the larger three dimensional metal organic framework (MOF) structure. We show that this material undergoes a two stage magnetic transition, related to the low dimensionality of the Mn lattice. First at 5 K, which is assigned to the initial development of short range order in the 2D layers. This is followed by long range order at 3.3 K. Applied field measurements reveal the potential to induce magnetic transitions in moderately small fields of $\sim$2 T. Neutron powder diffraction enabled the determination of a unique magnetic space group $P2_1'/c$ (\#14.77) at 1.5 K. This magnetic structure consists of antiferromagnetically coupled Mn-Mn dimers with spins principally along the out of plane $a$-axis. 
\end{abstract}

\maketitle

\section{Introduction}

Investigations of two-dimensional (2D) layered magnetic materials has revealed  properties of interest for both fundamental and applied research. This is exemplified by graphene and beyond-graphene materials where the bulk compound has 2D layers weakly coupled through van der Waals bonding that can often be exfoliated or otherwise isolated down to a few or single layers \cite{NatureGraphene, Novoselovaac9439, doi:10.1021/acs.chemrev.7b00633, Burch2018, doi.org/10.1038/s41563-020-0791-8, RevModPhys.90.015005}. In these inorganic compounds numerous non-trivial topological and quantum behaviors have been observed and predicted due to the low dimensionality enhancing this behavior. Investigations on inorganic quasi-2D materials have uncovered topologically protected Skyrmions \cite{WOS:000514255400009, PhysRevB.103.104410}, quantum spin liquids with emergent Majorana fermions \cite{Banerjee1055} and spontaneous topological Hall effect \cite{WOS:000972820500002}. Conversely, the tunability of coordination polymers, or equivalently magnetic metal-organic frameworks (MOFs), offers a powerful but less explored material space to achieve analogous physics when magnetic metal ions are added to well isolated 2D layered coordination structures \cite{ZHENG20141, PhysRevLett.108.077208,C7CS00653E, doi:10.1021/cr200304e,C8DT02411A, B804757J}. An extremely large variety of structures are available through often highly predictable organic chemistry routes. The ability to control the in-plane 2D motif, the spacing of layers, and potential introduce hybrid functionality on the organic linkers affords multiple intriguing research avenues for magnetic coordination polymers in the realm of quantum materials.

The material [Mn(pzc)$_2$]$_n$ (pzc = pyrazinecarboxylate), henceforth referred to as Mn-pyrazine, contains well isolated 2D layers of Mn$^{2+}$ ions. Ref.~\onlinecite{WOS:000179572700020} is the only previous literature report on this material, with x-ray diffraction and magnetic susceptibility measurements. The structure contains only one Mn site, however the Mn ions within the layer have two distinct bonding environments with spacings of $\sim$3.5 $\rm \AA$ and $\sim$5.6 $\rm \AA$. This results in Mn-Mn dimers that are linked to form a distorted 2D hexagonal network. These 2D layers are well isolated by an interlayer Mn-Mn distance of $\sim$9.4 $\rm \AA$, with the bonding containing weak hydrogen bonds between ligands. Mn-pyrazine is therefore expected to be a good realization of a quasi-2D material in a bulk compound. The previous powder magnetic susceptibility measurements were fit to a Curie-Weiss law down to 5 K and showed antiferromagnetic interactions with the Mn$^{2+}$ ion in the S=5/2 spin state. Inspecting the magnetic susceptibility in Ref.~\onlinecite{WOS:000179572700020} reveals an apparent low temperature anomaly, however there was no discussion of any potential magnetic order transition.

Here, we undertake  magnetic susceptibility, heat capacity and neutron powder diffraction measurements that reveal long range magnetic order in Mn-pyrazine. The directional dependence of the field behavior is investigated through single crystal magnetic susceptibility measurements that show a two stage magnetic transition. In addition there is a potential in-field magnetic transition, which indicates routes to tune the magnetic properties in moderate fields. Neutron powder diffraction is utilized to investigate the crystalline and magnetic structure through temperature dependent measurements. Despite the presence of increased incoherent scattering from hydrogen in Mn-pyrazine, empirical formula C$_{10}$H$_6$MnN$_4$O$_4$, good quality neutron diffraction data are obtained. This highlights the strength of monochromatic high flux reactor based neutron instruments coupled with the choice of large moment metal ions when investigating magnetic MOFs. At the lowest temperature of 2 K long range magnetic order is observed with several magnetic reflections identified. Symmetry analysis of this magnetic structure shows that the Mn-Mn dimers form antiferromagnetic pairs, with spins preferentially aligned in the out of plane $a$-direction in the $P2_1'/c$ $(\#14.77)$ magnetic space group.

\section{Methods}

\subsection{Synthesis}

Single crystals of Mn-pyrazine were grown using the low-temperature hydrothermal method. First, a total of 0.37 grams of pyrazinecarboxylate acid (C$_5$H$_4$N$_2$O$_2$) and anhydrous manganese(II) acetate (Mn(CH$_3$COO)$_2$) were mixed in a stoichiometric ratio of 2:1 with 5 mL of water and 5 mL of ethanol in a small beaker. The mixture was then mixed using a magnetic stirrer until everything was fully dissolved. Then the final mixture was loaded into a Teflon-lined stainless-steel autoclave, sealed well and heated at 140$^{\circ}$C for 12 hrs. After cooling to room-temperature a dark yellow solution was recovered and left to evaporate at room temperature. The Mn-pyrazine crystals were formed during the solvent evaporation. 

\subsection{Magnetic property characterization}
 
Temperature-dependent and field dependent magnetic measurements were performed using a Quantum Design Magnetic Property Measurement System (MPMS). Magnetic properties were determined using one single crystal with the weight of 3.8 mg. The single crystal specimen was affixed to a quartz rod using GE varnish and temperature dependent magnetization measurements were carried out along three crystal directions. The temperature dependent data were collected in the range of 2 - 350 K in an applied magnetic field up to 50 kOe. The anisotropic isothermal magnetization measurements were performed between 2 - 100 K in magnetic fields up to 60 kOe. The heat capacity ($\rm C_p$) of the sample was measured using a Physical Property Measurement System (PPMS) between 2 - 200 K under magnetic field in the range 0-60 kOe.

\subsection{Neutron powder diffraction}

Neutron powder diffraction measurements on 5 grams of undeuterated Mn-pyrazine were carried out on the HB-2A powder diffractometer at the High Flux Isotope Reactor (HFIR), Oak Ridge National Laboratory (ORNL) \cite{Garlea2010, doi:10.1063/1.5033906}. Hydrogen based materials present an extra challenge for neutron scattering by both adding to the neutron absorption and creating an increased background from incoherent scattering. These effects can be easier to account for with constant wavelength instruments due to the simpler data correction. Constant wavelength measurements were performed at 2.41 $\rm \AA$ from the Ge(113) monochromator reflection and 1.54 $\rm \AA$ from the Ge(115) reflection. The pre-mono, pre-sample and pre-detector collimation was open-open-12'. A pyrolytic graphite (PG) filter was placed before the sample to remove higher order reflections for the 2.41 $\rm \AA$ wavelength. The sample was contained in a 6 mm diameter vanadium can and cooled in a liquid $^4$He cryostat in the temperature range 1.5 K - 300 K. The diffraction pattern was collected by scanning a 120$^{\circ}$ bank of 44 $^3$He detectors in 0.05$^{\circ}$ steps to give 2$\theta$ coverage from 5$^{\circ}$ to 130$^{\circ}$. Counting times were 8 hours for the 2.41 $\rm \AA$ measurements and 2 hours for the 1.54 $\rm \AA$ wavelength.  Rietveld refinements were performed with Fullprof \cite{Fullprof}. Symmetry allowed magnetic structures were considered using both representational analysis with SARAh \cite{sarahwills} and magnetic space groups with the Bilbao Crystallographic Server \cite{Bilbao_Mag}. Plots of the crystal and magnetic structure were prepared using VESTA \cite{VESTA}.

\begin{figure}[tb]
	\centering         
	\includegraphics[trim=0.0cm 0cm 8cm 0cm,clip=true, width=0.85\columnwidth]{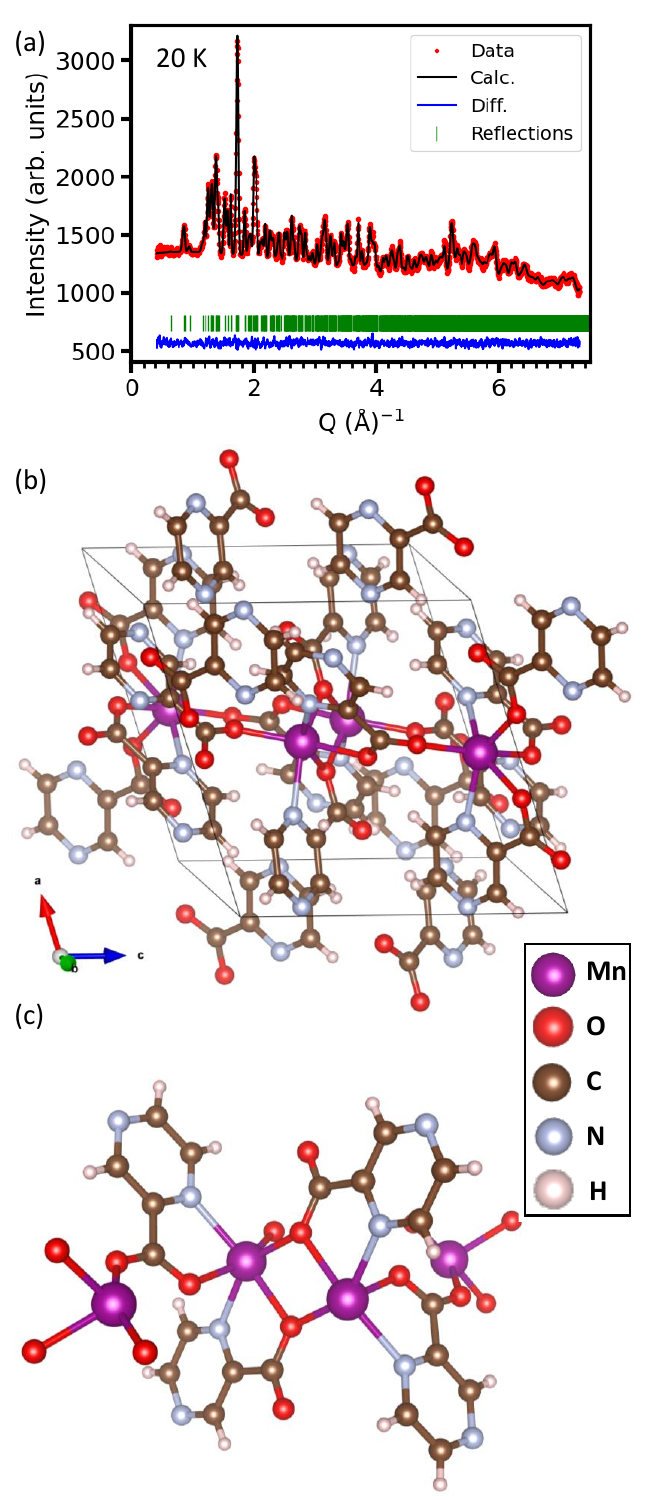}           
	\caption{\label{Fig_crystal} (a) Rietveld refinement of neutron powder diffraction data for Mn-pyrazine collected on the HB-2A instrument with a wavelength of 1.54 $\rm \AA$ at 20 K. (b) Crystal structure of Mn-pyrazine. The box represents the unit cell in the $P2_1/c$ space group. (c) Nearest and next nearest neighbor Mn ions are bonded through distinct pathways of Mn-O-Mn and Mn-O-C-O-Mn.}
\end{figure}

\begin{figure}[tb]
	\centering         
	\includegraphics[trim=0.0cm 2.5cm 3.8cm 0cm,clip=true, width=0.9\columnwidth]{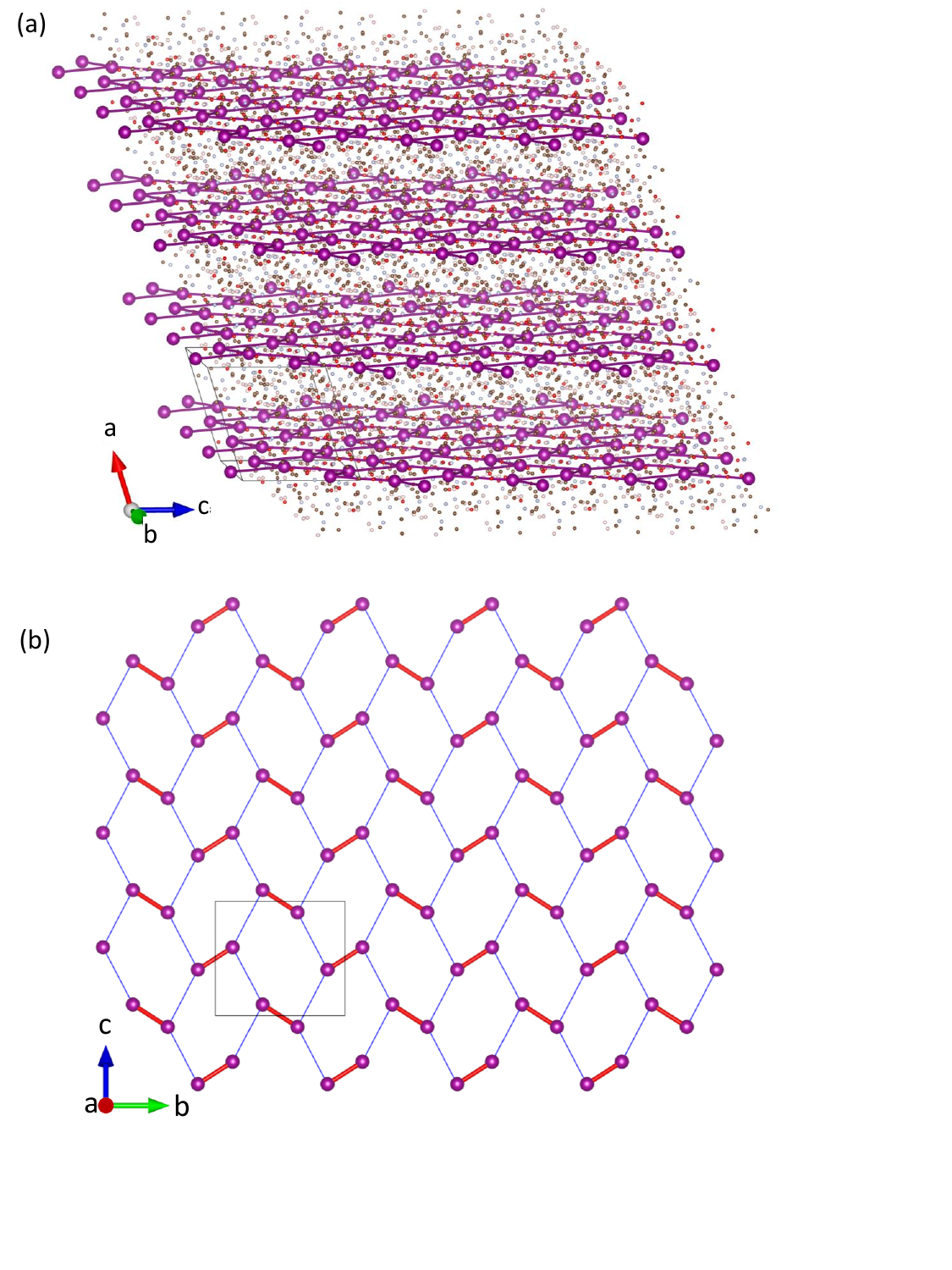}           
	\caption{\label{Fig_crystal_2Dlayers} (a) View of the 2D layers in Mn-pyrazine. (b) The nearest neighbor dimers (thick red line) and next nearest neighbor (thin blue line) interactions within the 2D layers are shown between the Mn ions (purple sphere). The Mn ions form a distorted 2D honeycomb structure. The crystallographic unit cell is indicated by the grey box.}
\end{figure}

\begin{table}[tb]
	\caption{\label{FPtableperamsMn}Refined crystal structure parameters for Mn-pyrazine, empirical formula C$_{10}$H$_6$MnN$_4$O$_4$, at 20 K for space group $P2_1/c$ with lattice constants a=10.2078(4) $\rm \AA$,  b=10.8444(4)  $\rm \AA$, c=10.1095(4) $\rm \AA$, $\alpha$=90$^{\circ}$, $\beta$=108.429(3)$^{\circ}$, $\gamma$=90$^{\circ}$.\\ 
		$\chi^2=  1.72$, $\rm R_{wp}$=8.80.}
	\begin{tabular}{c c c c c}
		\hline 
		Atom  &  $x$ & $y$& $z$ & site   \\ \hline
Mn  &  0.498(3) &   0.128(2) &   1.101(3) & $4e$   \\ 
C   &  0.835(2) &   0.119(1) &   1.146(1) & $4e$   \\ 
C   &  0.968(2) &   0.147(1) &   1.176(2) & $4e$   \\ 
H   &  1.038(3) &   0.104(2) &   1.130(2) & $4e$   \\ 
C   &  0.937(1) &   0.289(1) &   1.326(1) & $4e$   \\ 
H   &  0.974(3) &   0.368(2) &   1.394(2) & $4e$   \\ 
C   &  0.798(1) &   0.263(1) &   1.292(1) & $4e$   \\ 
H   &  0.727(2) &   0.305(2) &   1.351(2) & $4e$   \\ 
C   &  0.373(2) &   0.374(1) &   0.966(2) & $4e$   \\ 
C   &  0.271(1) &   0.468(1) &   0.918(1) & $4e$   \\ 
H   &  0.287(3) &   0.523(2) &   0.840(2) & $4e$   \\ 
C   &  0.159(1) &   0.417(1) &   1.079(2) & $4e$   \\ 
H   &  0.103(2) &   0.436(3) &   1.121(2) & $4e$   \\ 
C   &  0.258(2) &   0.322(1) &   1.125(1) & $4e$   \\ 
H   &  0.250(2) &   0.272(2) &   1.194(2) & $4e$   \\ 
C   &  0.756(1) &   0.018(1) &   1.035(1) & $4e$   \\ 
C   &  0.479(1) &   0.337(1) &   0.891(1) & $4e$   \\ 
N   &  0.740(1) &   0.176(1) &   1.205(1) & $4e$   \\ 
N   &  1.026(1) &   0.232(1) &   1.273(1) & $4e$   \\ 
N   &  0.360(1) &   0.297(1) &   1.066(1) & $4e$   \\ 
N   &  0.182(1) &   0.497(1) &   0.976(1) & $4e$   \\ 
O   &  0.829(2) &  -0.034(2) &   0.973(2) & $4e$   \\ 
O   &  0.627(2) &  -0.004(2) &   1.039(1) & $4e$   \\ 
O   &  0.551(2) &   0.247(1) &   0.939(1) & $4e$   \\ 
O   &  0.476(2) &   0.416(1) &   0.784(2) & $4e$   \\ 
		\hline
	\end{tabular}
\end{table}

\section{Results and Discussion}

\subsection{Crystal structure of Mn-pyrazine}

We begin by considering the crystal structure of Mn-pyrazine, which was previously reported as being in the $P2_1/c$ space group \cite{WOS:000179572700020}. To confirm this we carried out neutron powder diffraction measurements in the paramagnetic regime of 20 K. The shorter wavelength of 1.54 $\rm \AA$ was used to give the widest Q coverage to increase the number of measured reflections. The data has the expected elevated background and Q dependence from the incoherent hydrogen scattering. There are, however, well resolved and strong nuclear Bragg peaks. This data was refined with the reported $P2_1/c$ space group, with the background readily accounted for by a simple 6 coefficient polynomial function used for most measurements on this neutron powder diffractometer.  The data and refinement show good agreement, see Fig.~\Ref{Fig_crystal}(a). The refined lattice constants and atomic positions are given in Table~\Ref{FPtableperamsMn}. Due to the large number of parameters the thermal parameters were fixed during the analysis. 

The structural unit cell is shown in Fig.~\Ref{Fig_crystal}(b). Considering the local Mn environment in Fig.~\Ref{Fig_crystal}(c) indicates two distinct bonding pathways. Nearest neighbor bonds are Mn-O-Mn, which would suggest standard superexchange magnetic pathways. Whereas the next nearest neighbor bonds are Mn-O-C-O-Mn, requiring extended superexchange magnetic interactions. The nearest neighbor distance is 3.46(5) $\rm \AA$ and next nearest neighbor is 5.71(3) $\rm \AA$. The Mn-Mn distance between the layers is 10.2078(5) $\rm \AA$ and is mediated by a complex exchange pathway which includes weak hydrogen bonds. This creates the 2D layered structure of interest. Considering multiple unit cells, as shown in Fig.~\Ref{Fig_crystal_2Dlayers}(a), highlights the well isolated 2D network of Mn-Mn ions. The 2D network in the $bc$-plane is shown in Fig.~\Ref{Fig_crystal_2Dlayers}(b). The nearest neighbor Mn bonds can be viewed as Mn-Mn dimers which interact with the next nearest neighbor Mn ions to form a distorted 2D hexagonal layer.

\subsection{Magnetic susceptibility and heat capacity results}

\begin{figure}[tb]
	\centering         
	\includegraphics[trim=0.0cm 0.5cm 9.0cm 0cm,clip=true, width=0.76\columnwidth]{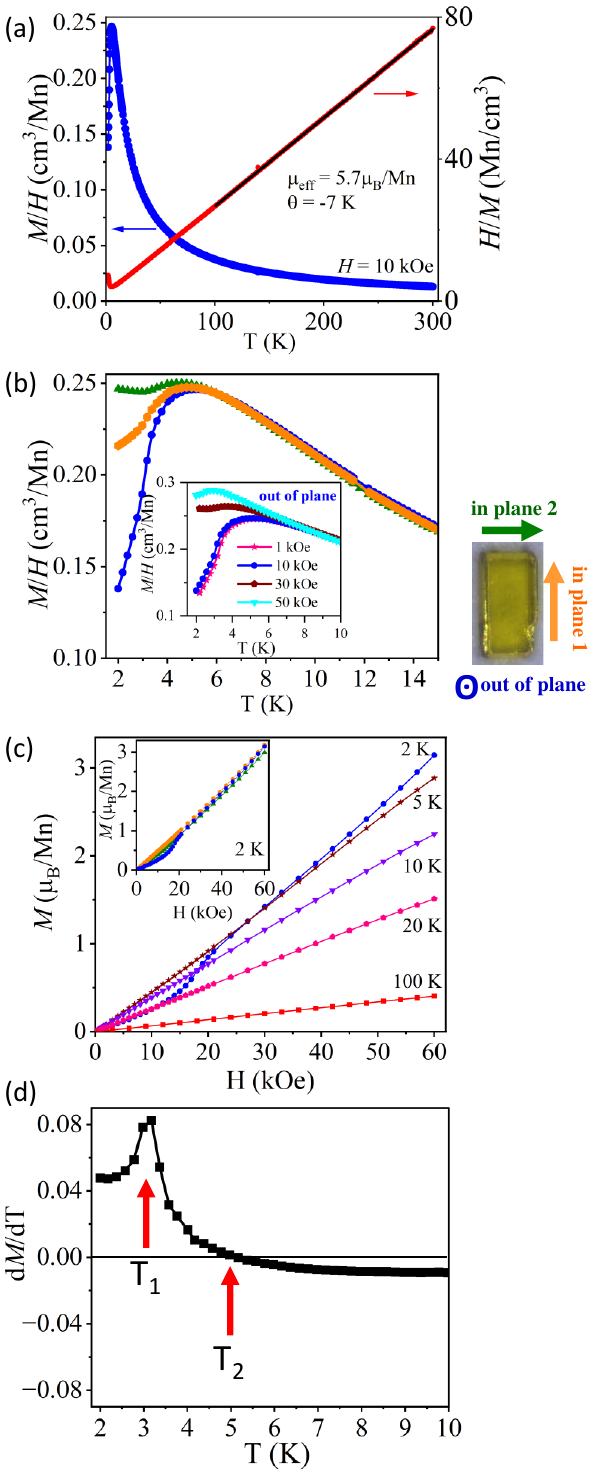}           
	\caption{\label{Fig_susceptibility} (a) Magnetic susceptibility measurements from 2 K to 300 K. The solid black line is a fit using the Curie-Weiss law. (b) Directional dependent measurements showing a broad anomaly centered around 5 K. Measurements were in a 10 kOe field for all three directions. (Inset) out of plane magnetic field dependence. (c) Isothermal magnetization measurements for a field applied out of the plane. (Inset) Directional dependent measurements at 2 K. (d) Derivative of the susceptibility reveals two anomalies at 5 K and 3.2 K in a field of 10 kOe out of the plane}
\end{figure} 

\begin{figure}[tb]
	\centering         
	\includegraphics[trim=0.0cm 0cm 2.5cm 0cm,clip=true, width=0.9\columnwidth]{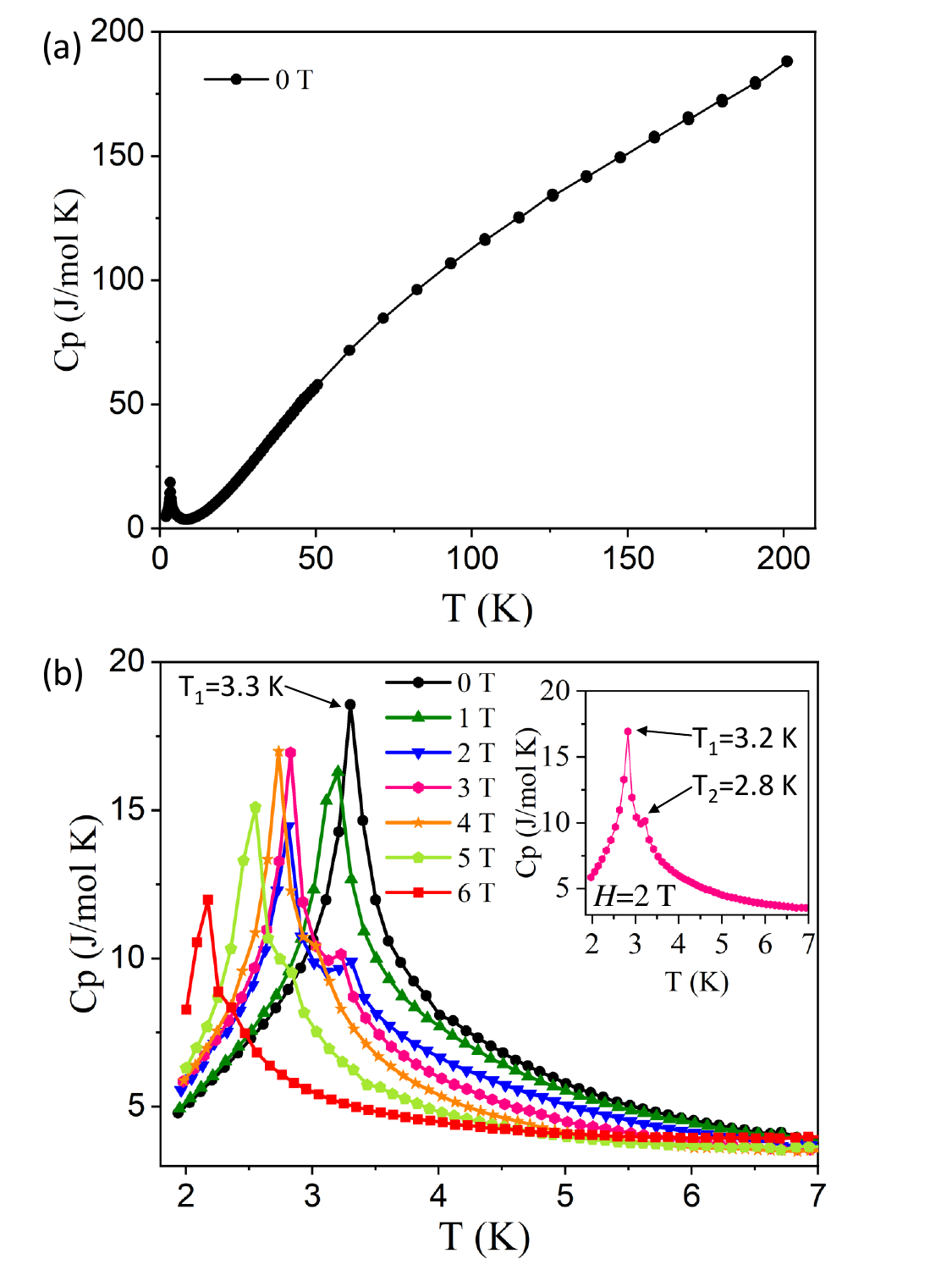}           
	\caption{\label{Fig_HeatCapacity} (a) Zero field heat capacity measurements on Mn-pyrazine from 2 to 200 K. (b) Low temperature and field dependent heat capacity measurements. }
\end{figure}

The initial report of Mn-pyrazine in Ref.~\onlinecite{WOS:000179572700020} measured magnetic susceptibility of a powder sample. Despite the apparent low temperature anomaly recorded, no detailed discussion of potential magnetic ordering was reported. In the synthesis reported here, small Mn-pyrazine crystals grew as rectangular rods. Therefore, we performed our magnetic measurements on single crystals in three orientations as displayed in Fig.~\Ref{Fig_susceptibility}. The magnetic susceptibility produces a broad peak below 10 K with a maximum around 5 K. After that, it continuously decreases down to the lowest temperature measured of 2 K, indicating an antiferromagnetic transition. A broad transition is often a signature of short-range ordering associate with the layered nature of the structure. The inverse susceptibility was fitted using the Curie-Weiss law in the range of temperature 100 $<$ T $<$ 300 K, see Fig.~\Ref{Fig_susceptibility}(a). This gives  $\theta_{\rm CW}$ = -7 K, indicating the antiferromagnetic nature of this compound. The effective magnetic moment from the fit was $\mu_{\rm eff}$ = 5.7 $\mu_B$/Mn, which is comparable to the expected moment of a Mn$^{2+}$ ion in the high-spin $d^5$ state of 5.9$\mu_B$.

The magnetic susceptibility measured in different crystal orientations are displayed in Fig.~\Ref{Fig_susceptibility}(b)-(c).  All field directions of in-plane and out-of-plane show a broad peak around the same temperature of 5 K at low fields $\le 10$ kOe. Increasing the field shifts the peak to lower temperatures. In general, the behavior under applied field is non-trivial with potentially a spin-flop transition which is further confirmed from our isothermal magnetization measurements in Fig.~\Ref{Fig_susceptibility}(c). This shows an anomaly at 2 K and 15 kOe. The inset of Fig.~\Ref{Fig_susceptibility}(c) showing directional dependent measurements indicates this anomaly occurs for fields applied out of the plane only. Considering the derivative of susceptibility (d$M$/dT) indicates a two-stage transition at T$_2$= 5 K and T$_1$=3 K.

To gain further insights into the low temperature behavior in Mn-pyrazine we performed heat capacity measurements. Figure \Ref{Fig_HeatCapacity} shows the results. The peak in the heat capacity for 0 T is observed at 3.3 K. This is lower then the broad peak from magnetic susceptibility which occurs around 5 K, but consistent with the T$_2$ transition in the d$M$/dT analysis. Applying a field leads to a lowering of the transition temperature in the heat capacity measurements and the observation of a further transition in the form of a shoulder in the heat capacity anomaly. This can be seen in the inset of Fig.~\Ref{Fig_HeatCapacity} for a field of 3 T where a shoulder is measured at 3.2 K, followed by a further anomaly at 2.8 K. 

Collectively the susceptibility and heat capacity results reveal multiple transitions that can be rationalized by considering the low dimensional nature of Mn-Pyrazine. We postulate that short range order occurs around 5 K in the 2D layers. This gives the broad peak in the susceptibility, rather than a sharp transition. Then at 3.3 K the transition to three dimensional long range order occurs. This is observed in the heat capacity and also the magnetic susceptibility through d$M$/dT as a two stage transition.

\subsection{Magnetic structure of Mn-pyrazine}

\begin{figure}[tb]
	\centering         
	\includegraphics[trim=0.0cm 2cm 0cm 0cm,clip=true, width=1.0\columnwidth]{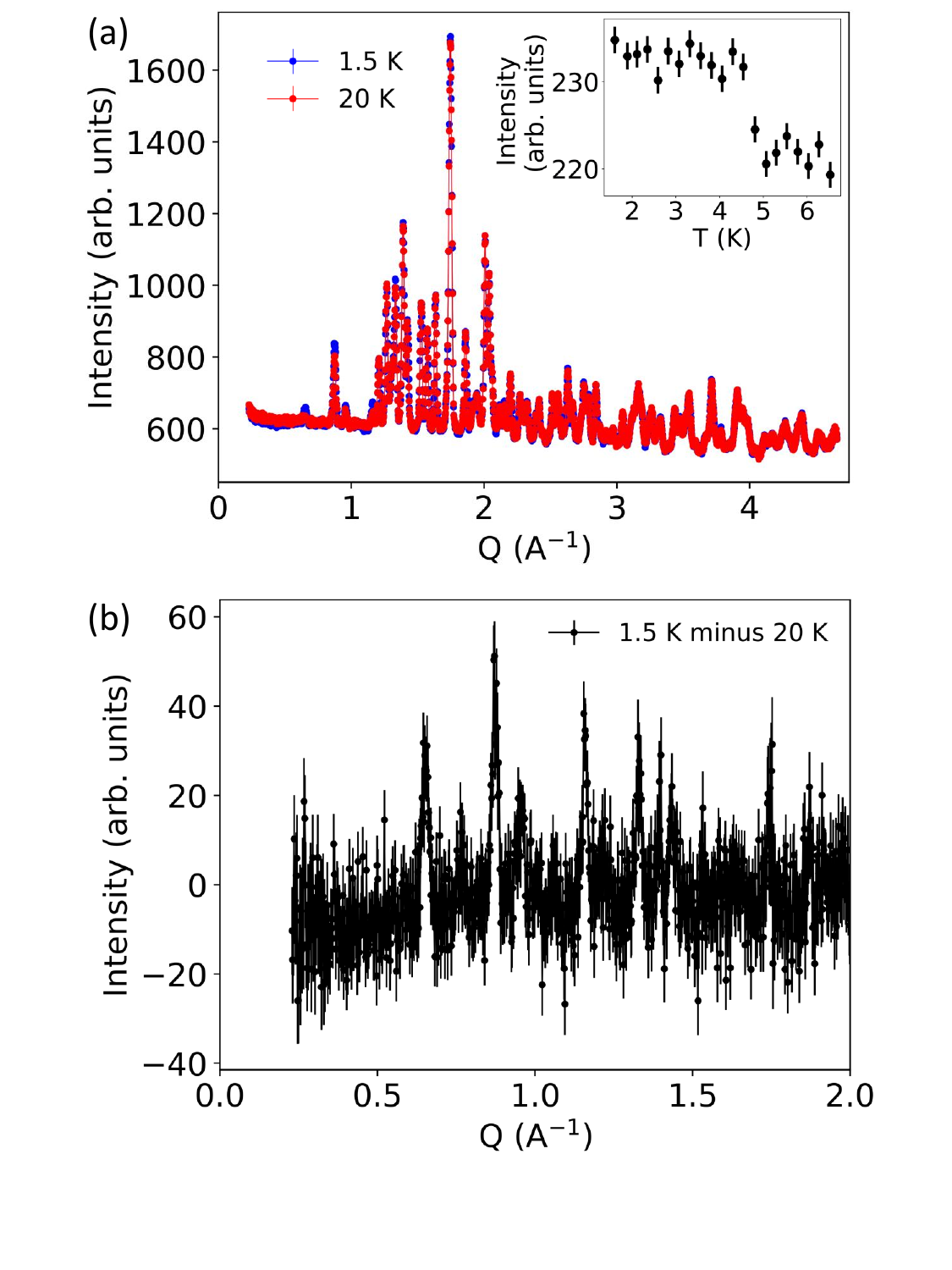}           
	\caption{\label{Fig_HB2A_data} (a) Neutron powder diffraction data collected at 1.5 K and 20 K. (Inset) Intensity at Q=0.65$\rm \AA$ as a function of temperature through the magnetic transition. (b) Difference of intensity at 1.5 K and 20 K in the powder diffraction data.}
\end{figure} 

\begin{figure*}[tb]
	\centering         
	\includegraphics[trim=0.0cm 4cm 0cm 0cm,clip=true, width=1.0\textwidth]{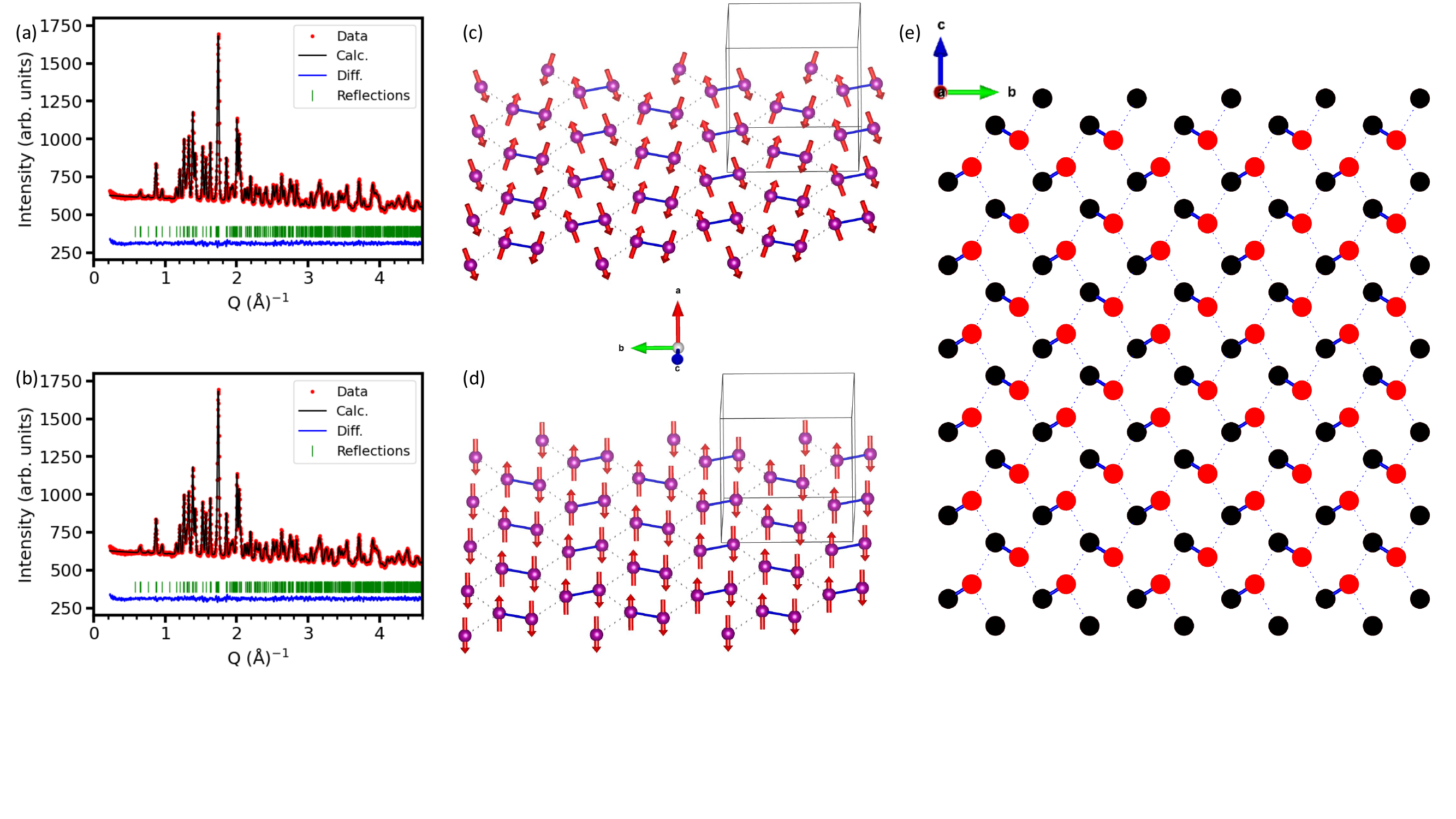}           
	\caption{\label{Fig_MnPyr_HB2A_FP_MagneticRefinements}  (a) Refinement of neutron powder diffraction data collected at 1.5 K using the $P2_1'/c$ with moments along all ($\bf{m}_a$,$\bf{m}_b$,$\bf{m}_c$) directions. (b) Refinement of neutron powder diffraction data collected at 1.5 K using the $P2_1'/c$ with moments only along $\bf{m}_a$. Magnetic structures for (c) ($\bf{m}_a$,$\bf{m}_b$,$\bf{m}_c$) and (d) ($\bf{m}_a$,0, 0) models. The moments are shown as red arrows on the Mn ions, the nearest neighbor dimer bonds are blue and the next nearest neighbor bond are the dashed lines. The grey box represents the magnetic unit cell. (e) View of the 2D layers along the $a$-axis. The red/black circles correspond to up/down spin directions on the Mn ion.}
\end{figure*} 

\begin{table}[tb]
	\begin{tabular}{cc|cll}
		IR  &  BV  &  Point Group & \multicolumn{2}{c}{Magnetic Space Group} \\
		\hline
		$\Gamma_{1}$ & $\psi_{1}$ &    $2/m$ & $P2_1/c$ & $(\#14.75)$  \\
		& $\psi_{2}$ &    $2/m$ & $P2_1/c$ & $(\#14.75)$  \\
		& $\psi_{3}$ &    $2/m$ & $P2_1/c$ & $(\#14.75)$  \\
		$\Gamma_{2}$ & $\psi_{4}$ &    $2/m$ & $P2_1/c'$ & $(\#14.78)$  \\
		& $\psi_{5}$ &    $2/m$ & $P2_1/c'$ & $(\#14.78)$  \\
		& $\psi_{6}$ &    $2/m$ & $P2_1/c'$ & $(\#14.78)$  \\
		$\Gamma_{3}$ & $\psi_{7}$ &    $2/m$ & $P2_1'/c'$ & $(\#14.79)$  \\
		& $\psi_{8}$ &    $2/m$ & $P2_1'/c'$ & $(\#14.79)$  \\
		& $\psi_{9}$ &    $2/m$ & $P2_1'/c'$ & $(\#14.79)$  \\
		$\Gamma_{4}$ & $\psi_{10}$ &    $2/m$ & $P2_1'/c$ & $(\#14.77)$  \\
		& $\psi_{11}$ &    $2/m$ & $P2_1'/c$ & $(\#14.77)$  \\
		& $\psi_{12}$ &    $2/m$ & $P2_1'/c$ & $(\#14.77)$  \\
	\end{tabular}
	\caption{The point symmetry and magnetic space group for the space group $P2_1/c$ with ${\bf k}=(0,0,0)$. The decomposition of the magnetic representation for the Mn site $(0.510,0.134,0.598)$ is $\Gamma_{Mag}=3\Gamma_{1}^{1}+3\Gamma_{2}^{1}+3\Gamma_{3}^{1}+3\Gamma_{4}^{1}$.}
	\label{bv_symmetry_vector_table_1}
\end{table}

\begin{table}[h]
	\begin{tabular}{ccc|ccc}
		IR  &  BV  &  Atom & \multicolumn{3}{c}{BV components}\\
		&      &             &$m_{\|a}$ & $m_{\|b}$ & $m_{\|c}$  \\
		\hline
		$\Gamma_{4}$ & $\psi_{10}$ &      1 &      1 &      0 &      0   \\
		&              &      2 &      1 &      0 &      0  \\
		&              &      3 &     -1 &      0 &      0  \\
		&              &      4 &     -1 &      0 &      0   \\
		& $\psi_{11}$ &      1 &      0 &      1 &      0   \\
		&              &      2 &      0 &     -1 &      0   \\
		&              &      3 &      0 &     -1 &      0  \\
		&              &      4 &      0 &      1 &      0   \\
		& $\psi_{12}$ &      1 &      0 &      0 &      1   \\
		&              &      2 &      0 &      0 &      1   \\
		&              &      3 &      0 &      0 &     -1   \\
		&              &      4 &      0 &      0 &     -1   \\
	\end{tabular}
	\caption{Basis vectors for the space group $P2_1/c$ with ${\bf k}=(0,0,0)$. The decomposition of the magnetic representation for the Mn site $(0.51009,~0.13426,~0.598)$ is $\Gamma_{Mag}=3\Gamma_{1}^{1}+3\Gamma_{2}^{1}+3\Gamma_{3}^{1}+3\Gamma_{4}^{1}$. The atoms of the nonprimitive basis are defined according to 1:~$(0.510,0.134,0.598)$, 2:~$(0.490,0.634,0.902)$, 3:~$(0.490,0.866,.402)$, 4:~$(0.510,0.366,0.098)$.}
	\label{basis_vector_table_1}
\end{table}

We now turn to neutron powder diffraction measurements to investigate the microscopic magnetic spin structure. Diffraction patterns were collected at 1.5 K and 20 K using the 2.41 $\rm \AA$ wavelength, see Fig.~\Ref{Fig_HB2A_data}(a). The high Q scattering is unchanged, indicating no structural symmetry change. The low Q behavior below 2 $\rm \AA^{-1}$, however, reveals additional intensity in the form of Bragg peaks. These can be assigned to magnetic ordering, with the width of the new magnetic peaks the same as the nuclear peaks, indicating long range magnetic order at 1.5 K. The change in scattering is emphasized in the temperature difference plot in Fig.~\Ref{Fig_HB2A_data}(b). 

The intensity of the peak at Q=0.65 $\rm \AA^{-1}$, which has no observable nuclear contribution at 20 K, was measured as a function of temperature to follow the onset of magnetic ordering. Increased scattering is observed at 5 K, which corresponds to the broad peak in magnetic susceptibility. This increases, until it becomes saturated below 3 K. This measurement was exclusively of the intensity at Q=0.65 $\rm \AA^{-1}$. It did not allow for a measurement the peak width, therefore it is not possible to distinguish between short or long range order scattering. Since any diffuse scattering intensity will be small, it would be of interest for future studies to measure deuterated powder or larger single crystals to further investigate the potential short range order in Mn-pyrazine through this transition.

At 1.5 K Mn-pyrazine is in the long ranged magnetically ordered state. The measured magnetic Bragg peaks can all be indexed with a ${\bf k}$=(0,0,0) propagation vector. Starting from the paramagnetic space group of $P2_1/c$ and using the determined propagation vector gives equivalently four irreducible representations (IRs) in a representational analysis approach or four maximal magnetic space groups. The magnetic space groups are $P2_1/c$ (\#14.75), $P2_1'/c$ (\#14.77), $P2_1/c'$ (\#14.78) and $P2_1'/c'$ (\#14.79). The corresponding IRs $\Gamma_1$, $\Gamma_2$, $\Gamma_3$ and $\Gamma_4$, in Kovalevs representation, are shown in Table.~\Ref{bv_symmetry_vector_table_1}.

For all candidate models there is symmetry allowed spin components along all crystallographic directions  ($\bf{m}_a$,$\bf{m}_b$,$\bf{m}_c$). The data was refined against all 4 candidate magnetic models. Only magnetic space group $P2_1'/c$ $(\#14.77)$ ($\Gamma_4$) was able to reproduce the intensity of all observed magnetic reflections. Allowing the moments to refine freely gave ($\bf{m}_a$,$\bf{m}_b$,$\bf{m}_c$)= (4.363(101),  1.261(562),  1.532(237) ) and total moment 4.3(2)$\mu_B$/Mn$^{2+}$. Which is reduced but nevertheless close to the full spin for a S=5/2 ion of 5$\mu_B$. This magnetic structure is shown in Fig.\Ref{Fig_MnPyr_HB2A_FP_MagneticRefinements}(c). Confining the spins to only have a component along the $a$-axis gives an equivalently good refinement, as can be seen in Fig.\Ref{Fig_MnPyr_HB2A_FP_MagneticRefinements}(b). The corresponding moment is slightly increased and closer to the full S=5/2 values with ($m_a$,$m_b$,$m_c$)=  (4.47(9),0,0)  and total moment 4.47(9) $\mu_B$/Mn$^{2+}$. This magnetic structure is shown in Fig.\Ref{Fig_MnPyr_HB2A_FP_MagneticRefinements}(d). Allowing the spins to only be constrained to either the $b$ or $c$ axis could not fully account for the data. These results confirm the S=5/2 nature of the Mn ion, however the reduction in moment may be a consequence of the extended nature of the Mn-Mn exchange pathways leading to a degree of moment delocalization onto the surrounding ligands.

The spin behavior in the 2D layer can be visualized in Fig.\Ref{Fig_MnPyr_HB2A_FP_MagneticRefinements}(e). The black/red correspond to up/down Mn moments. The magnetic structure of Mn-pyrazine has antiferromagnetic dimers of Mn-Mn ions. Each next nearest neighbor Mn-Mn interaction in the layer is also antiferromagnetic. Despite the large Mn-Mn distance of $>$ 10 $\rm \AA$ three dimensional long range order occurs, with the nearest neighbor Mn-Mn interlayer correlation being ferromagnetic. The large Mn moment of S=5/2 is likely a driving factor in realizing long range order. 

To induce further interesting behavior in Mn-pyrazine and related materials it will be of interest to reduce the moment size down to S=1/2 to enhance quantum phenomena. This will be of particular interest with the dimer and 2D hexagonal layers that are hosts to exotic physics. As one example the Mn-pyrazine structure can be considered a distorted Shastry-Sutherland lattice \cite{PhysRevLett.82.3701, doi:10.1073/pnas.1413318111}. Removing or controlling this distortion may provide routes to investigate this physics in MOFs and allow for a wider phase space of materials than the currently limited candidates.

\section{Conclusions}

Mn-pyrazine (C$_{10}$H$_6$MnN$_4$O$_4$) has been investigated with magnetic susceptibility, heat capacity and neutron powder diffraction. The magnetic susceptibility and heat capacity collectively indicate development of short range order at 5 K that proceeds the long range magnetic phase transition at 3.2 K in zero field. The applied field measurements show anisotropic behavior with a field driven anomaly above 2 T. Moreover heat capacity measurements in fields above 2 T reveal two observable anomalies with a small temperature window of $\sim$0.5 K. Neutron powder diffraction was able to determine the magnetic structure in this undeuterated material. Following a symmetry analysis only a single magnetic space group,  $P2_1'/c$ (\#14.77), was found to be consistent with the data. The analysis revealed the moments primarily aligned along the $a$-axis, which is out of the 2D layers. These moments form antiferromagnetic dimers that are linked within a wider distorted hexagonal network.  In general the results show that coordination polymers, or equivalently magnetic metal-organic frameworks (MOFs), with both organic and inorganic building blocks offer unique material avenues to explore tailored structural motifs due to the versatility and predictability of organic chemistry.

\section{acknowledgments}
This research used resources at the High Flux Isotope Reactor, a DOE Office of Science User Facility operated by the Oak Ridge National Laboratory. This research used resources at the Missouri University Research Reactor (MURR). This work was supported in part by a University of Missouri Research Council Grant (Grant Number: URC-22-021). This manuscript has been authored by UT-Battelle, LLC under Contract No. DE-AC05-00OR22725 with the U.S. Department of Energy. The United States Government retains and the publisher, by accepting the article for publication, acknowledges that the United States Government retains a non-exclusive, paidup, irrevocable, world-wide license to publish or reproduce the published form of this manuscript, or allow others to do so, for United States Government purposes. The Department of Energy will provide public access to these results of federally sponsored research in accordance with the DOE Public Access Plan(http://energy.gov/downloads/doepublic-access-plan).

\end{document}